\documentclass[a4paper]{article}

\usepackage[english]{babel}
\usepackage[utf8x]{inputenc}
\usepackage[T1]{fontenc}
\usepackage{helvet}

\usepackage{upgreek} 
\usepackage{soul} 
\usepackage{lineno} 

\usepackage[a4paper,top=3cm,bottom=2cm,left=3cm,right=3cm,marginparwidth=1.75cm]{geometry}
\renewcommand{\arraystretch}{1.3}
\newcommand{\beq}{\begin{equation}}
\newcommand{\eeq}{\end{equation}}

\usepackage{amsmath}
\usepackage{graphicx}
\usepackage[colorinlistoftodos]{todonotes}
\usepackage[colorlinks=true, allcolors=blue]{hyperref}
\usepackage{array}
\usepackage{float}
\usepackage{appendix} 
\usepackage{multirow}
\usepackage[symbol]{footmisc}
\usepackage{setspace}
\usepackage{authblk} 
\usepackage{ctable}

\usepackage{comment}

\usepackage{cite}
\usepackage{amsmath,amssymb,amsfonts}
\usepackage{algorithmic}
\usepackage{graphicx}
\usepackage{textcomp}
\usepackage{caption}
\usepackage[square,numbers,,sort&compress]{natbib}

\begin{document}

This work has been submitted to the IEEE for possible publication. Copyright may be transferred without notice, after which this version may no longer be accessible.

\clearpage
\title{An Engineer's Brief Introduction to Microwave Quantum Optics and a Single-Port State-Space Representation}
\author[1]{Malida O. Hecht}
\author[1]{Antonio J. Cobarrubia}
\author[1]{Kyle M. Sundqvist}

\affil[1]{Department of Physics, San Diego State University, San Diego, CA 92182}
\date{}
\maketitle

\begin{abstract}
Classical microwave circuit theory is incapable of representing some phenomena at the quantum level.  To include quantum statistical effects when treating microwave networks, various theoretical treatments can be employed such as quantum input-output network (QION) theory and \textit{SLH} theory. However, these require a reformulation of classical microwave theory. To make these topics comprehensible to an electrical engineer, we demonstrate some underpinnings of microwave quantum optics in terms of microwave engineering. For instance, we equate traveling-wave phasors in a transmission line ($V_0^+$) directly to bosonic field operators.  Furthermore, we extend QION to include a state-space representation and a transfer function for a single port quantum network. This serves as a case study to highlight how microwave methodologies can be applied in open quantum systems. Although the same conclusion could be found from a full \textit{SLH} theory treatment, our method was derived directly from first principles of QION. 
\end{abstract}

\section{Introduction}
\label{sec:introduction}
Although classical circuit theory has been the forefront of microwave engineering for many decades,  the theory is incapable of explaining noise from single-photon detection, long distance communication, phase amplifiers, etc \citep{Haus_1995}. With growing interest to utilize microwave networks for quantum communication \citep{Haus_1995}, quantum computing \citep{Devoret_2014},  quantum information \citep{Devoret2013SuperconductingCF},  and quantum networking \citep{Combes_2017}, a second quantization of circuit components has provided a successful transition from classical to quantum treatment for microwave networks. Such models include quantum input-output network (QION) theory \citep{Gardiner_1985,Gardiner_1992,Gardiner_2004,Yurke_1984,Drummond_2004} and \textit{SLH}  theory \citep{Gough_2008,Kockum_2012,Bouten_2019}, which \textit{SLH} has success in applying classical methodologies to quantum mechanics including a state-space representation. 

The difficultly of these models arises from the need to accurately describe the dynamics of an open quantum system. In particular, a popular treatment in both optical and microwave networks is by expressing bosonic operators in terms of an quantum input/output operators (Gaussian noise increments). From here, a quantum stochastic differential equation can fully describe the dynamics of a quantum network \citep{Gardiner_1985}. This is the fundamental principle behind QION and \textit{SLH} theory, where \textit{SLH} expands on this idea to include multiport networks. This makes QION theory a useful tool to solve simple single-port quantum networks, however, in literature, QION formalism was dominated by the perspective of quantum optics with no clear transition for microwave theory. This becomes a concerning problem when describing microwave systems using QION since the research field diverges from standardized notation depending on one's interest. In addition, there has been little attempt to connect QION to classical methodologie, which can potentially expand already established microwave techniques to quantum networks. 


This paper was developed to outline the importance of microwave formalism for a simple single port circuit connected to a transmission line.  In Section \ref{sec:Classical_normal_modes_of_the_lumped_LC_oscillator}, we introduced classical normal modes of a lumped LC oscillator to later compare classical methodology to quantum circuit theory. In Section \ref{sec:Quantization_of_electric_circuits}, we derived quantum LC operators of nodal charge and nodal flux in the form of annihilation operators. We formulated a comprehensible definition of a bosonic wave operator(s) for a 1-dimensional transmission line, and showed how to map between classical microwave formalism to quantum electric circuit theory. Lastly,  in Section \ref{sec:heisenberg_equation_of_motion_for_quantum_networks} we apply these definitions to create a state-space representation and transfer function for a single-port lumped LC oscillator coupled to a transmission line. 

Although the methodologies applied in this article can be used to solve nonlinear networks such as Josephson junctions and qubits \citep{Krantz_2019,Gu_2017}, the approximations made in nonlinear circuits do not affect the overall treatment of QION compared to a linear network. To illustrate the analogies between classical and quantum methodologies, this article focuses on a simple single-port lumped LC example. This approach can be easily extended to a multiport system and/or more complicated systems.

\section{Classical Normal Modes of the Lumped LC Oscillator}
\label{sec:Classical_normal_modes_of_the_lumped_LC_oscillator}
We first consider a lumped LC oscillator, consisting of an inductor with inductance $L$ and capacitor with capacitance $C$. The equations of motion for LC circuit are given by \citep{louisell_1960}
\begin{equation}
    \begin{aligned}
        \dfrac{{dI}}{{dt}} &= -\dfrac{1}{ {L}} {V}  \\ 
        \dfrac{ {dV}}{ {dt}} &= \dfrac{1}{ {C}} {I} .
    \end{aligned}
    \label{eq:LumpedEOM}
\end{equation}
which, through linear combinations, yields the normal mode equations:
\begin{equation}
    \begin{aligned}
        a &= \dfrac{1}{2}\sqrt{ {L}} ( {I} +i\omega  {CV})  \\
         a^* &= \dfrac{1}{2}\sqrt{ {L}} ( {I} -i\omega  {CV}) 
    \end{aligned}
    \label{eq:LCnorm}
\end{equation}
where $i$ is the imaginary unit and $\omega$ is the natural oscillation frequency $\omega = \frac{1}{\sqrt{ {LC}}}$. The normalization provides that the square of the amplitudes represent the energy stored in the modes, and the sum of the squares of the modes give the total energy of the system,
$
     {E} =\lvert a(t)\rvert^2+ \lvert a^*(t)\rvert^2= \frac{1}{2} \left[  {CV}^2(t) + {LI}^2(t)\right].
$

This understanding of normal modes in the classical sense helps us understand the form necessary to introduce bosonic operators.
\section{Quantization of Electric Circuits} 
\label{sec:Quantization_of_electric_circuits}
\subsection{Quantization of the Lumped LC Oscillator}
\label{sec:sub:Quantization_of_the_lumped_LC_oscillator}
To quantize the lumped LC circuit, we can follow the same conventions from \citep{louisell_1960}, starting with the same equations of motion as Eq. (\ref{eq:LumpedEOM}). However, since most quantum engineering systems involve Josephson Junctions, it will be more convenient to consider the equations of motion in terms of node flux and charge. We define node flux as $\Phi(t) = \int^t dt'V(t')$, such that voltage is given by $V(t) = \frac{d\Phi}{dt}$, we can also define curent to be $I(t) =- \frac{\Phi}{L}(t)$. Using the fact that node charge is given by $Q(t) = C V(t) $, we can rewrite Eq. (\ref{eq:LumpedEOM}) in terms of flux and charge:
\begin{equation}
      \begin{aligned}
   \dfrac{d\Phi}{dt}& = \dfrac{Q}{C} \\ 
       \dfrac{dQ}{dt} &= -\dfrac{\Phi}{L}.
    \end{aligned}
    \label{eq:QLumpedEOM}
\end{equation}
 This same result can be obtained by finding the Lagrangian and Hamiltonian of the system, where flux $\Phi$ is the analog to displacement, and charge $Q$ is its momentum conjugate. They can then be made quantum operators that follow the canonical commutation relation:
 
\begin{equation}
   [\hat{\Phi},\hat{Q}] = i \hbar . 
\end{equation}

We find the quantized normal mode operators to be \citep{girvin_leshouches,Clerk_2010}: 
\begin{equation}
\begin{aligned}
\hat{a} = \cfrac{\hat{\Phi}}{\sqrt{2L\hbar\omega}} + i\cfrac{\hat{Q}}{\sqrt{2C\hbar \omega}}, \\
\hat{a}^\dagger = \cfrac{\hat{\Phi}}{\sqrt{2L\hbar\omega}} - i\cfrac{\hat{Q}}{\sqrt{2C\hbar \omega}}.
\end{aligned}
 \label{eq:LCnormQ}
\end{equation}
Where $\omega$ is the resonant frequency given by $\omega = \frac{1}{\sqrt{LC}}$. Substituting $\Phi = LI$ and $Q = CV$ into these quantized normal modes returns Eq. (\ref{eq:LCnorm}) normalized by $(\sqrt{2\hbar\omega})^{-1}$, which ensures that the commutation relation $[\hat{a},\hat{a}^\dagger] = 1$ holds, and that our operators are dimensionless, rather than having units of square root energy, like the classical normal modes of the lumped oscillator.  

By rearranging Eq. (\ref{eq:LCnormQ}), one can write flux, charge, current and voltage in terms of bosonic operators $\hat{a}$ and $\hat{a}^\dagger$
\begin{equation}
    \begin{aligned}
    \hat{\Phi} &=L \hat{I}= \sqrt{\dfrac{L\hbar\omega}{2}}(\hat{a} +\hat{a}^\dagger),\\
    \hat{Q} &= C \hat{V}= -i\sqrt{\dfrac{C\hbar\omega}{2}}(\hat{a}-\hat{a}^\dagger), 
    \end{aligned}
\end{equation}
Additionally, the Hamiltonian is given by $\mathcal{H} = \hbar\omega(\hat{a}^\dagger \hat{a} + \frac{1}{2})$

\subsection{Quantization of the Lossless Transmission Line}
\label{sec_sub:Quantization_of_the_lossless_transmission_line}
We now turn to consideration of the lossless transmission line, beginning with the familiar Telegrapher's equations \citep{louisell_1990,pozar,ulaby,Raju}
\begin{equation}
    \begin{aligned}
    \dfrac{\partial }{\partial z}V(z,t) &= -L'\dfrac{\partial }{\partial t}I (z,t)\\
    \dfrac{\partial }{\partial z}I (z,t) &= -C' \dfrac{\partial }{\partial t}V (z,t),
    \end{aligned}
    \label{eq:telegraphers}
\end{equation}
where $L'$ and $C'$ are inductance and capacitance per unit length, respectively. Again, like with the quantization of the lumped oscillator, it will be convenient to consider the system in terms of node flux and node charge. We relate voltage as $ \frac{\partial}{\partial t}\Phi(z,t) = V(z,t) $ and $I = -\frac{1}{L'}\frac{\partial }{\partial z}\Phi(z,t)$. Then, substituting our relations, we end up getting 1-d wave equations in terms of flux
\begin{equation}v_p^2  \dfrac{\partial^2 }{\partial z^2} \Phi(z,t)= \dfrac{\partial^2}{\partial t^2}\Phi(z,t), \label{eq:1Dwaveeq}\end{equation}
where $v_p$ is the phase velocity given by $v_p = \frac{1}{\sqrt{L'C'}}$
We note that Eq. (\ref{eq:1Dwaveeq}) is the same result obtained by considering the Lagrangian of the system, and finding the Euler-Lagrange equation of motion, where node flux is the analog to the displacement done by Refs. \citep{girvin_leshouches,BeltranThesis,Clerk_2010, Drummond_2004}. To remain consistent, we will consider solutions to Eq. (\ref{eq:1Dwaveeq}). It can be shown that the charge density, $q$, (charge per unit length) is the conjugate momentum to flux, with the commutation relation given by
 \begin{equation}
   [\hat{\Phi}(z,t),\hat{q}(z',t)] = i \hbar \delta(z-z'). 
\label{eq:commphi} \end{equation}

We can then present normal modes for the transmission line given by \citep{girvin_leshouches}:
 \begin{equation}
     \begin{aligned}
     \hat{b}_k =\sqrt{ \dfrac{\omega_k C'\ell}{2\hbar}}\hat{\Phi}_k+ i \sqrt{\dfrac{1}{2C'\ell\hbar\omega_k}}\hat{Q}_k \\
     \hat{b}_k^\dagger =\sqrt{ \dfrac{\omega_k C'\ell}{2\hbar}}\hat{\Phi}_k- i \sqrt{\dfrac{1}{2C'\ell\hbar\omega_k}}\hat{Q_k},
     \end{aligned} \label{eq:bose}
 \end{equation}
 where $\ell$ is a length along the transmission line, typically where the transmission line becomes periodic, $k$ is the wave number and $\omega_k = v_p k$. 
It can be shown that the commutation relation is given by $[\hat{b}_k , \hat{b}_k^\dagger] = 1$, ensured by the normalization. $\hat{b}_k$ and $ \hat{b}_k^\dagger$ are the familiar Bose lowering and raising operators, respectively. The Hamiltonian is then given the familiar form of $\mathcal{H}  = \sum_k \hbar\omega_k \left( \hat{b}_k^\dagger\hat{b}_k + \frac{1}{2}\right)$ for a quantized simple harmonic oscillator. Equation (\ref{eq:bose}) yields the node flux and node charge at a specified mode $k$ given by
 \begin{equation}
     \begin{aligned}
     \hat{\Phi}_k &= \sqrt{\dfrac{\hbar}{2\omega_k C'\ell}}\left(\hat{b}_k + \hat{b}_k^\dagger\right) \\
     \hat{Q}_k &= -i \sqrt{\dfrac{C'\ell\hbar\omega_k}{2}}\left(\hat{b}_k - \hat{b}_k^\dagger\right)
     \end{aligned}
 \end{equation}
and they can can be considered as travelling waves of flux and charge, respectively \citep{BeltranThesis,Clerk_2010}
 \begin{equation}
     \begin{aligned}
     & \hat{\Phi}_R(z,t) =  \\
     & \ \ \ \ \ \sum_{k>0} \sqrt{\dfrac{\hbar}{2\omega_k C'\ell}}\left(\hat{b}_ke^{-i(\omega_k t - kz)} + \hat{b}_k^\dagger e^{i (\omega_k t - kz)}\right) \\
     & \hat{Q}_R(t) =  \\
     & \ \ \ \ \ -i \sum_{k>0} \sqrt{\dfrac{C'\ell\hbar\omega_k }{2}}\left(\hat{b}_ke^{-i(\omega_kt-kz)} - \hat{b}_k^\dagger e^{i( \omega_k t-kz)}\right).
     \end{aligned}
     \label{eq:fluxchargewave}
 \end{equation}
Using the fact that $Q = C'\ell V$ and $-L'\ell I = \Phi$, or by considering time derivatives of Eq. (\ref{eq:fluxchargewave}), one can find the right-moving voltage and current waves to be 
\begin{equation}
\begin{aligned}
&\hat{V}_R (z,t) =  \\
& \ \ \ \ \ -i \sum_{k>0} \sqrt{\frac{\hbar\omega_k}{2C'\ell}}\left(\hat{b}_k e^{-i(\omega_k t-kz)}-\hat{b}_k^\dagger e^{i(\omega_k t-kz)}\right) \\
&\hat{I}_R (z,t) =  \\
& \ \ \ \ \ \dfrac{-1}{L'\ell} \sum_{k>0} \sqrt{\dfrac{\hbar}{2\omega_k C'\ell}}\left(\hat{b}_k e^{-i(\omega_k t-kz)}+\hat{b}_k^\dagger e^{i(\omega_k t-kz)}\right)
\end{aligned}
\label{eq:voltagesum}
\end{equation}
It can be shown that $\left[\hat{I}_R, \hat{V}_R  \right] \neq 0$, and similarly for the left-moving voltage and current waves, which contributes to zero point fluctuations in power. We notice that voltage and current, described in  Eq. (\ref{eq:voltagesum}) is offset by a phase $\phi = \frac{\pi}{2}$ in comparison to microwave theory. This is due to the fact that the choice of origin is arbitrary and we chose flux to be our position coordinate, and charge to be momentum. To match notation in microwave theory, we can introduce a phase offset of $\pi /2$ which maps $\hat{b'}_k =\hat{b}_k e^{i\pi/2}= i\hat{b}_k$ and $\hat{b'}_k^\dagger=\hat{b}_k^\dagger e^{-i\pi/2}= -i\hat{b}_k^\dagger$. Substituting this into Eq. (\ref{eq:voltagesum}), we obtain voltage as a sum of the Bose operators that is real:

\begin{equation}
    \begin{aligned}
     & \hat{V}_R (z,t)   =    \\
    & \sum_{k>0} \sqrt{\frac{\hbar\omega_k}{2C'\ell}} \left( \hat{b}'_k e^{-i(\omega_k t-kz)} + \hat{b}_k'^\dagger e^{i(\omega_k t-kz)} \right).
    \end{aligned}
    \label{eq:voltagereal}
\end{equation}
Similarly, current is given by
\begin{equation}
\begin{aligned}
    & \hat{I}_R (z,t) = \\
    & \frac{i}{L'\ell} \sum_{k>0} \sqrt{\dfrac{\hbar}{2\omega_k C'\ell}}\left(\hat{b'}_k^\dagger e^{i(\omega_k t-kz)}-\hat{b'}_k e^{-i(\omega_k t-kz)}\right).
    \end{aligned}
\end{equation}

Following references \citep{Clerk_2010,Drummond_2004,BeltranThesis}, we can extend Eq. (\ref{eq:voltagereal}) to the continuum limit, where $\ell \rightarrow \infty$,
\begin{equation}
    \begin{aligned}
    \hat{V}_R(z,t) = \int_0^\infty \dfrac{d\omega}{2\pi}\sqrt{\dfrac{\hbar \omega Z_0}{2}}\left( \hat{b}_R (\omega) e^{-i(\omega t-kz)}+h.c.\right)\\
    \hat{V}_L(z,t) =\int_0^\infty \dfrac{d\omega}{2\pi}\sqrt{\dfrac{\hbar \omega Z_0}{2}}\left( \hat{b}_L (\omega) e^{-i(\omega t+kz)}+h.c. \right), \label{eq:voltint}
    \end{aligned}
\end{equation}
where $Z_0 $ is the characteristic impedance given by $Z_0 = \sqrt{\frac{L'}{C'}}$, and
\begin{equation}
\begin{aligned}
\hat{b}_R (\omega) = 2\pi (L'C'\ell^2)^{-1/4}\sum_{k>0}\hat{b'}_k \delta(\omega - \omega_k),  \\
\hat{b}_L (\omega) = 2\pi (L'C'\ell^2)^{-1/4}\sum_{k<0}\hat{b'}_k \delta(\omega - \omega_k). \label{eq:frequency representation}
\end{aligned}
\end{equation}
 The resulting commutation relation is given by
\begin{equation}
    \left[ \hat{b}_R(\omega),\hat{b}_R^\dagger(\omega') \right] = \left[ \hat{b}_L(\omega),\hat{b}_L^\dagger(\omega') \right] = 2\pi\delta(\omega - \omega') \label{eq:commomega}
\end{equation}
The normalization of Eq. (\ref{eq:frequency representation}) ensures that the commutation relations of Eq. (\ref{eq:commphi}) and Eq. (\ref{eq:commomega}) hold. Equation (\ref{eq:commphi}) is important because it depicts the uncertainty relationship between charge and flux, which are canonical conjugates. Equation (\ref{eq:commomega}) is important because we recover boson commutation relations. 

\subsection{Mapping Quantum Circuits by way of Microwave Engineering}
We now consider a classical, single-mode voltage traveling-wave, moving to the right, with instantaneous solution $V_R(z,t) = \left| V_0^+\right| \cos(\omega t-kz +\phi)$. Using sinusoidal steady-state analysis to provide phasor notation, ${V_R}\left( t \right) = \mathrm{Re} \left\{ {V_0^ + \exp \left[ {i\left( {\omega t - kz + \phi } \right)} \right]} \right\}$. Note that the Bose operators could also be written as  $\hat{b}_k= e^{-i\hat{\phi}}\sqrt{\hat{N}_k}$ and $\hat{b}_k^\dagger = \sqrt{\hat{N}_k}e^{i\hat{\phi}}$ \citep{zagoskin_2011}, where  $\hat{\phi}$ is a \textit{hypothetical} phase operator, defined uniquely \textit{only} for the interval $[0,2\pi]$ \citep{Carruthers_1968, TANAS1996355}.
We also note that $\hat{b}_k\hat{b}_k^\dagger = \hat{N}_k+1$ and $\hat{b}_k^\dagger \hat{b}_k  = \hat{N}_k$, which ensures that the commutation relation $[\hat{b}_k , \hat{b}_k^\dagger] = 1$ holds. We can consider these quantum-mechanical expressions in a semi-classical limit, where we propose bosonic operators now to be downgraded to the role of simple scalars.  In this limit, we can equate our prior expression for traveling-wave voltage with a standard, plane-wave engineering expression.  In this case, we recover the following relationships to standard microwave expressions \citep{ulaby,pozar,Raju},
\begin{equation}
\begin{aligned}
V_0^+ = \left| V_0^+\right| e^{i\phi}& = \sqrt{\dfrac{2\hbar\omega_k}{C'\ell}}\hat{b}_k^\dagger e^{-i\pi/2}, \ (k>0)\\
\hat{b}_k^\dagger = \sqrt{\hat{N}}e^{i\hat{\phi}}&= \sqrt{\dfrac{C'\ell}{2\hbar\omega_k}}V_0^+ e^{i\pi/2}, \ (k>0) .
\end{aligned}
\end{equation}
A relationship between $V_0^-$ and $\hat{b}_k^\dagger$ can be found by considering the left-moving voltages instead.
Using the relationship between $V_0^+$ and $\hat{b}_k^\dagger$, one can relate Pozar's formalism for generalized scattering parameters, $a = \frac{ V_0^+}{\sqrt{Z_0}}$ with $\hat{b}_k^\dagger$. A comprehensive translation -- possibly an interdisciplinary Rosetta Stone -- between formalisms can be seen in Table \ref{tab:mapping}.

It is tempting to speculate how one could upgrade parameters $V_0^+$ and $V_0 ^- $ to quantum operators,  $\hat{V}_0^+$ and $\hat{V}_0 ^- $, and derive open quantum systems in terms of these operators, rather than with using traditional bosonic operators.

\begin{table*}[h]
    \centering
     \renewcommand{\arraystretch}{3}
    \begin{tabular}{c|c}
      Microwave Engineering   &Quantum Engineering  \\ \hline
       $ V_0^+ $  & $\sqrt{\dfrac{2\hbar\omega_k}{C'\ell}}\hat{b}_k^\dagger e^{-i\pi/2}, \ (k>0)$\\
       $V_0^-$ &  $\sqrt{\dfrac{2\hbar\omega_k}{C'\ell}}\hat{b}_k^\dagger e^{-i\pi/2}, \ (k<0)$\\
       $\sqrt{\dfrac{C'\ell}{2\hbar\omega_k}}V_0^+ e^{i\pi/2}$ & $\hat{b}_k^\dagger$ \\
       $a= \dfrac{V_0^+}{\sqrt{Z_0}}$ & $\sqrt{\dfrac{2\hbar\omega_k v_p}{\ell}}\hat{b}_k^\dagger e^{-i\pi/2}, \ (k>0)$ \\
       $b= \dfrac{V_0^-}{\sqrt{Z_0}}$& $\sqrt{\dfrac{2\hbar\omega_k v_p}{\ell}}\hat{b}_k^\dagger e^{-i\pi/2},  \ (k<0)$\\
    \end{tabular}
    \caption{Quantum engineering Rosetta Stone of generalized scattering parameters in microwave engineering language \citep{pozar}, with bosonic operators in a quantized transmission line. The characteristic impedance is given by $Z_0 = \sqrt{\frac{L'}{C'}}$ and phase velocity given by $v_p= \frac{1}{\sqrt{L'C'}}$.
    \label{tab:mapping}}
\end{table*}
\section{Heisenberg Equation of Motion, Input-output relation and state-space representation}
\label{sec:heisenberg_equation_of_motion_for_quantum_networks}

\begin{figure}[h]
    \centering
    \includegraphics[width=4in,height=2in,clip,keepaspectratio]{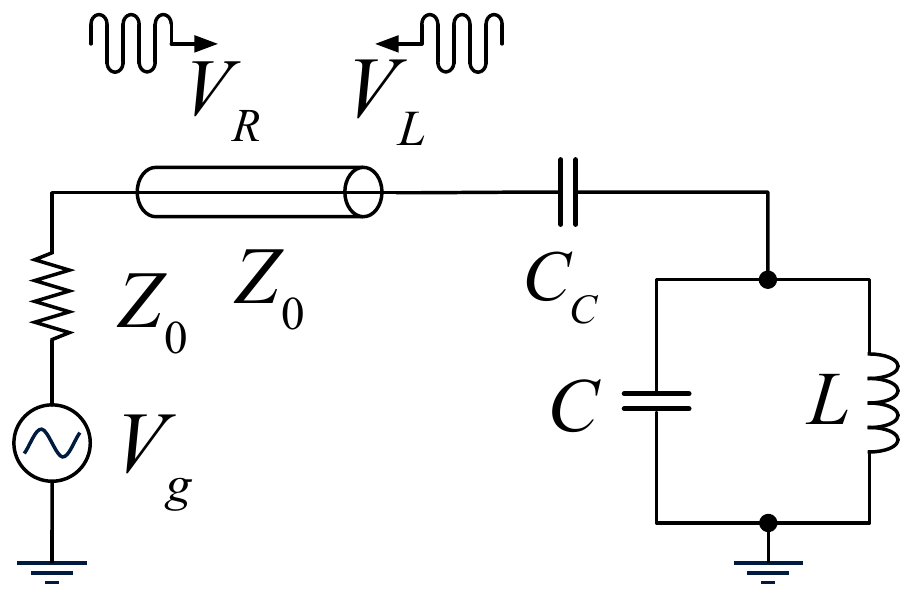}
    \caption{A lumped LC parallel tank circuit is coupled to a transmission line through a coupling capacitor.  Although thermal photons in a semi-infinite transmission line could also be considered, here we implicitly assume a generator to be matched at the source of the transmission line}
    \label{fig1}
\end{figure}

Assume a transmission line coupled to a quantum lumped LC oscillator (system) by a coupling capacitor with capacitance $C_c$, see Fig. \ref{fig1}. (The capacitance and inductance of the system are denoted as $L$ and $C$, respectively). This creates an open quantum system,
\begin{align}
    H = H_{sys} + H_{int} + H_{bath},
\end{align}
 where $H_{sys}$ is the Hamiltonian of lumped LC, $H_{int}$ is the Hamiltonian of the interaction between the lumped LC and bosonic bath operator (voltage wave), and $H_{bath}$ is the Hamiltonian of the bosonic field. We assume that the transmission line is semi-infinite where, at $z = 0$ and t = 0, the right-propagating wave $\hat{V}_R$ enters the lumped LC oscillator at t = 0,  and at a later time t the left propagating wave $\hat{V}_L$ is reflected back. This results in the bath Hamiltonian $H_{bath}$ defined in Section \ref{sec_sub:Quantization_of_the_lossless_transmission_line} where the right-moving bosonic operators are the input bosonic operators $b_{in} = \hat{b}_{R}(\omega)$ and, similarly, the output bosonic operator are the left-moving bosonic operators. The system Hamiltonian $H_{sys}$ is the Hamiltonian of the lumped LC oscillator described in Section \ref{sec:sub:Quantization_of_the_lumped_LC_oscillator} using the values in Fig. \ref{fig1} (right side). 

We assume the interaction Hamiltonian between the input voltage and system, $H_{int}$, to be the same as the energy stored in the coupling capacitor, $E_{C_c}$, where $H_{int} = E_{C_c} =  \frac{1}{2}C_c (\hat{V}_{bath} - \hat{V})^2$ (with C $\gg C_c$). We assume that $\frac{1}{2}C_c\hat{V}^2_{bath}$ and $\frac{1}{2}C_c\hat{V}^2$ are negligible compared to the bath and system Hamiltonian terms. This is because the impedance across the coupling capacitor will be much greater than the impedance across the system capacitor C. Therefore more energy will flow across the system capacitor than the coupling capacitor ($H \gg  \frac{1}{2}C_cV^2$). A similar argument can be used for $\frac{1}{2}C_c\hat{V}^2_{bath}$. Then interaction Hamiltonian reduces to $H_{int} \approx -C_{c}\hat{V}_{bath} \hat{V}$. 

Each Hamiltonian can be defined as
\begin{subequations}
\begin{align}
    H_{sys} & = \hbar \triangle a^\dagger a ,\\
    H_{int} & = i \frac{\hbar C_c}{4\pi}\sqrt{\frac{Z_0 \omega_R}{C}} \int_{0}^\infty \sqrt{\omega} (\hat{b}(\omega) + \hat{b}^\dagger (\omega)) (a - a^\dagger ) d\omega ,\\
    H_{bath} & = \hbar \int_0^\infty \omega \hat{b}^\dagger (\omega)\hat{b}(\omega)d\omega,
\end{align}
\end{subequations}

\noindent where $\triangle = \omega - \omega_R$ is the detuning frequency of the lumped LC, the resonant frequency for the circuit is $\omega_R = (LC)^{-\frac{1}{2}}$, $\hat{V}$ is the voltage operator in Section \ref{sec:sub:Quantization_of_the_lumped_LC_oscillator},   $Z_0$ is the characteristic impedance of the transmission line, and an implied tensor product between ($b(\omega)$ + $b^\dagger(\omega)$) and (a - $a^\dagger$). We also denote the identity matrix in the bosonic operator Hilbert space as $\mathbf{I}_{bath}$, and the identity matrix in the system's Hilbert space as $\mathbf{I}_{sys}$.

Before we solve the dynamics of this system, we would want to change the basis of the interaction Hamiltonian to interaction (Dirac) picture in the perspective of the system \citep{Gardiner_1985}. This is done by considering a bare Hamiltonian $H_0 = H_{bath} + H_{int}$ with $H_{int}(t) =e^{\frac{i}{\hbar}H_0 t}H_{int}e^{-\frac{i}{\hbar}H_0 t}$ such that 
\begin{align}
    H_{int}(t)  & = i\hbar \frac{C_c}{4\pi}\sqrt{\frac{Z_0 \omega_R}{C}} \int_{0}^\infty \sqrt{\omega} (\hat{b}(\omega) \nonumber \\
    & \ \ \ \ \ \ \ \ \ \ \ \ \ \ \ \ \ \ + \hat{b}^\dagger (\omega)) (a(t) - a^\dagger (t))d\omega
\end{align}
where a(t) = $ae^{-i\triangle t}$. Commutation relations still hold $[a(t),a^\dagger (t')] = \delta(t - t')$ and $[a^\dagger (t),a^\dagger (t')] = [a(t),a (t')] = 0$. In addition, the system operator in the interaction frame is modified to $H_{sys} = \hbar \triangle a^\dagger (t) a(t)$.

Let's consider  a wave at frequency $\Omega$. At large frequencies away from $\Omega$, the field has very little interaction with the system states \citep{Clerk_2010}. In consequence, we can integrate both the bath and interaction Hamiltonian on the interval $(-\infty, \infty$). This means for both the interaction and bath Hamiltonian the limits of integration can change as $\int_0^\infty \longrightarrow \int_{-\infty}^{\infty}$. 

For weak coupling between the system and bath, the interaction Hamiltonian as it is now doesn't conserve energy with counter-revolution terms are non-conservative $\hat{b}^\dagger (\omega) a^\dagger (t)$ and $\hat{b}(\omega) a(t)$ \citep{Gardiner_2004}. To fix this, we introduce the Rotating Wave Approximation (RWA) to get rid of these terms in the interaction Hamiltonian \citep{Loudon_2000,Gardiner_1985,Gardiner_1992,Gardiner_2004,Combes_2017}.

We assume that the rate of change in frequency is minuscule around $\Omega$ such that $\omega \approx \Omega $. This means we can make Markovian approximation $\frac{C_c}{4\pi}(\frac{Z_0 \omega_R}{C})^{\frac{1}{2}} \sqrt{\Omega} \approx \frac{\kappa}{\sqrt{2\pi}}$, where $\kappa$ is the constant coupling strength around resonance \citep{Gardiner_1985,Gardiner_1992,Gardiner_2004}. This approximation, $\kappa$, describes how strong bosons interact with the network for every unit $(\text{time})^{-\frac{1}{2}}$. The interaction Hamiltonian yields
\begin{align}
    H_{int}(t)  = i\hbar \frac{\kappa}{\sqrt{2\pi}} \int_{-\infty}^\infty  (\hat{b}^\dagger (\omega) a(t) - a^\dagger (t) \hat{b}(\omega) )d\omega,
\end{align} 
\noindent where with implied outer products between a(t) and $\hat{b}(\omega)$ operators. This modifies the bath Hamiltonian as $H_{bath} = \hbar \int_{-\infty}^\infty (\omega - \Omega) \hat{b}^\dagger (\omega)\hat{b}(\omega)d\omega$ \citep{Combes_2017}. Often the Markovian approximation is written as $\kappa(\omega) \approx \sqrt{\frac{\gamma}{2\pi}}$, which is the normal formalism in quantum optics as it connects $\gamma$ with how much photons are entering the network \citep{Abdo_2013} (units of $(\text{time})^{-1}$). However, we feel this nomenclature not to be as insightful in this derivation since coupling between bath and system goes by units of $(\text{time})^{-\frac{1}{2}}$, where the meaning of the dimensional unit analysis is lost. Therefore, we chose to highlight the significance of the "constant" coupling strength $\kappa$.

\subsection{Heisenberg-Langevin Equation}
\label{sec:sub:heisenberg_langevin_equation}

We want to solve the equation of motion for the bosonic field operators, specifically we want to develop an input bosonic operator. To do this we will have to solve the the Heisenberg equation of motion in the bosonic field frame. In the interaction frame of the bosonic field, the dynamics only depend on the interaction and bath Hamiltonian such that $\Dot{b}(\omega) = -\frac{i}{\hbar}[b(\omega), H_{int} + H_{bath}]$ to describes all the dynamics of $b(\omega)$ from $t_0$ to $t$. The solutions to this equation yields 
\begin{align}
    \hat{b}(\omega) & = \hat{b}(\omega,t = t_0)e^{-i (\omega -\Omega)(t- t_0)} \nonumber \\
    & \ \ \ \ \ \ + \frac{1}{\sqrt{2\pi}}\int_{t_0}^t \kappa a(t') e^{-i  (\omega -\Omega) (t'- t)} dt'. \label{eq:bosonic_heinsenberg_part}
\end{align}

We can find the evolution of the system operator a(t) from $t_0$ to $t$ in the Heisenberg picture (and interaction picture) just as we did with $\hat{b}(\omega)$. This equation is also known as the Heisenberg-Langevin equation \citep{Drummond_2004,Gardiner_1985,Gardiner_1992,Combes_2017} is given as 
\begin{align}
    \dot{a}(t) & = -\frac{i}{\hbar}[a(t),H_{sys} + H_{int}], \\
    & =  -i[a(t), \triangle a^\dagger (t) a(t)]  \nonumber \\
    & \ \ \ + [a(t), \frac{\kappa}{\sqrt{2\pi}} \int_{-\infty}^\infty  (\hat{b}^\dagger (\omega) a(t) - a^\dagger (t) \hat{b}(\omega) )d\omega ] , \label{eq:Heisenberg_equation_part1}
\end{align}
where there is an implied inner product with identity matrix of the bosonic field $(a(t) = a(t)\otimes \mathbf{I}_{bath})$.

Here we assume that the system state cannot evolve until the right propagating wave enters the network and, for a finite field between the times $t$ and $t + dt$, the field interacts with the system changing its state. In a similar manner, the wave reflected from the lumped LC cannot change (or be created) until the system state changes. This means that the bosonic field operator are independent (commutes) with system operator at integrated time $t'$ bigger the present time t. Therefore, $[\hat{b}(\omega),a(t')] = [\hat{b}^\dagger (\omega),a(t')] = [\hat{b}(\omega),a^\dagger(t')] = [b^\dagger (\omega),a^\dagger(t')] = 0$ when t' $> t$. Equation Eq. (\ref{eq:Heisenberg_equation_part1}) yields
\begin{align}
    \dot{a}(t) & =  -i \triangle a(t)  \nonumber \\
    & \ \ \ + \frac{\kappa}{\sqrt{2\pi}} \int_{-\infty}^\infty \hat{b}^\dagger (\omega) [a(t),  a(t)] d\omega \nonumber \\
    & \ \ \ \ - \frac{\kappa}{\sqrt{2\pi}} \int_{-\infty}^\infty  [a(t), a^\dagger (t) ] \hat{b}(\omega) d\omega 
    \end{align}
Plug in the commutation relations,
\begin{align}
    \dot{a}(t) & =  -i \triangle a(t) \nonumber \\
    &  \ \ \ \  \ \ \ \ \ \ \ \  \ \ \ \ - \frac{\kappa}{\sqrt{2\pi}} \int_{-\infty}^\infty  \hat{b}(\omega) d\omega\label{eq:Heisenberg_equation_part2}
\end{align}

We define the bosonic field (at $z = 0$) entering the lumped LC at $t_0 = 0$ as the input bosonic field 
\begin{align}
    b_{in}(t) & = \frac{1}{\sqrt{2\pi}} \int_{-\infty}^\infty \hat{b}(\omega,t = 0) e^{-i   (\omega -\Omega) t} d\omega,
\end{align}
where $\hat{b}(\omega,t = 0)$ is initial field entering the system in frequency domain. The output operator is derived in a similar manner, but with $t_0 = t_1$ where $t_1$ is the first "infinitesimal slice" of the wave is changed  by the system state. 

Plugging Eq. (\ref{eq:bosonic_heinsenberg_part}) into Eq. (\ref{eq:Heisenberg_equation_part2}),
\begin{align}
    \dot{a}(t) & =  -i  \triangle a(t) - \kappa \big( b_{in}(t) \nonumber \\
    & \ \ \ \ \ \ + \frac{1}{2\pi} \int_{-\infty}^\infty \int_{0}^t \kappa a(t') e^{-i  (\omega -\Omega) (t'- t)} dt' \big)
\end{align}

Using properties in \citep{Gardiner_1985} where $\frac{1}{2\pi}\int_{-\infty}^{\infty} e^{-i  (\omega -\Omega) (t -t')} d\omega = \delta(t - t')$ and 2 $\int_{t_0}^{t} \delta(t - t')a(t')dt' = sgn(t-t_o)a(t) $ as we integrate between $t$ and $t + dt$. For this network we have solved the Heisenberg-Langevin equation based on input-output theory \citep{Gardiner_1985,Drummond_2004}: 
\begin{align}
    \dot{a}(t) & = - (i\triangle  + \frac{\kappa^2}{2}) a(t) - \kappa b_{in}(t), \label{eq:heisenberg_langevin}
\end{align}
where a(0) = a $\otimes \mathbf{I}_{bath}$.

\subsection{Input-output relation}
\label{sec:sub:input_output_relation}
The most important result in input-output theory is the map between an input field to its output field. Reference \citep{Gardiner_1985} derived a general form of this based on the negligible time delay between the input port and output port. Using this assumption, the solution of Eq. (\ref{eq:bosonic_heinsenberg_part}) for $t_0 = 0$ (the input field) equals the solution of the wave function for the output field ($t_0 = t_1$) at some $\tau$. This yields
\begin{align}
  &  \hat{b}(\omega,t = 0)e^{-i  (\omega -\Omega) t}  \nonumber \\
  & \ \ \ \ \ \ \ \ \ \ \ \ \ \ \ \ \ \ \ \ \ \ \ \ + \frac{1}{\sqrt{2\pi}}\int_{0}^\tau \kappa a(t') e^{-i  (\omega -\Omega) (t'- t)} dt' \nonumber \\
  & = b(\omega,t = t_1)e^{-i  (\omega -\Omega) (t- t_1)} \nonumber \\
 & \ \ \ \ \ \ \ \ \ \ \ \ \ \ \ \ \ \ \ \ \ \ \ \ + \frac{1}{\sqrt{2\pi}}\int_{t_1}^\tau \kappa a(t') e^{-i  (\omega -\Omega) (t'- t)} dt'. \label{eq:input_output_relation_p1}
\end{align}
Integrating from $\omega$ to $\omega + d\omega$ in Eq. (\ref{eq:input_output_relation_p1}) on both sides gives, 
\begin{align}
    \sqrt{2\pi}b_{in}(t) + \frac{1}{\sqrt{2\pi}} \int_0^\tau \kappa a(t') 2\pi \delta(t - t') dt' \nonumber \\
     = \sqrt{2\pi}b_{out}(t) - \frac{1}{\sqrt{2\pi}} \int_{\tau}^{t_1} \kappa a(t') 2\pi \delta(t - t') dt' \\
     \sqrt{2\pi}b_{in}(t) + \sqrt{2\pi} \frac{\kappa}{2} a(t) = \sqrt{2\pi}b_{out}(t) - \sqrt{2\pi} \frac{\kappa}{2} a(t) \label{eq:input_output_relation_p2}
\end{align}
Divide Eq. (\ref{eq:input_output_relation_p2}) by $\sqrt{2 \pi}$ and rearrange the equation so $b_{out}(t)$ is by itself on the left hand side. This yields the input-output relation (originally derived by Gardiner and Collett \citep{Gardiner_1985})
\begin{align}
    b_{out}(t) & = \kappa a(t) + b_{in}(t), \label{eq:input_output_relation}
\end{align}
where $b_{in/out}(t)$ is denoted as $\mathbf{I}_{sys} \otimes b_{in/out}(t)$ and $a(t)$ as $a(t) \otimes \mathbf{I}_{bath}$.
\subsection{State-Space Representation}
\label{sec:sub:ABCD_representation}

The equations of motion derived from the Heisenberg-Langevin equation and Input-Output theorem can be combined into a system of equations; also known as a state-space representation if the equations describe the entire dynamics of the network \citep{Combes_2017, Tabak_2016,Mabuchi_2008,Devoret_1997QuantumFI,Ogata_2010,Williams_2007}. The system operator $a(t)$ is a ladder operator with quadratic degrees of freedom, which can be treated as a state variable. The bosonic field operators are input and output vectors. This means solutions to the Heisenberg-Langevin equation and input-output relation can fully describe the dynamics of a quantum network. Here we can connect the dynamics to a state-space representation, 
\begin{subequations}
\begin{align}
    \dot{a}(t) & = \mathbf{A} \ a(t) + \mathbf{B} \ b_{in}(t), \label{eq:ABCD_part_1}\\
    b_{out}(t) & = \mathbf{C} \ a(t) + \mathbf{D} \ b_{in}(t). \label{eq:ABCD_part_2}
\end{align}
\end{subequations}
Since the solutions to Eq. (\ref{eq:heisenberg_langevin}) and Eq. (\ref{eq:input_output_relation}) can completely describe the dynamics of the LC system operator $a(t)$ and the reflected (output) wave $b_{out}(t)$ given an input operator $b_{in}(t)$, then \textbf{A} = -(i$\triangle + \frac{\kappa^2}{2})$, \textbf{B} = -$\kappa$, \textbf{C} = $\kappa$, and \textbf{D} = 1. In this example, the \textit{ABCD} matrices are 1 $\times$ 1 dimensional since this is a 1-port network which correspond to the number of the ports for the quantum LC oscillator.

As in classical state-space representation, the \textit{ABCD} formulation for linear quantum systems is the dynamical equation of the system operator and the output wave function from 0 to $t$. The  lumped LC circuit described in  Eq. (\ref{eq:ABCD_part_1})- (\ref{eq:ABCD_part_2}) is an appropriate description of a passive network since the matrix of coefficients --\textit{ABCD} matrices--  only scale for a non-hermitian conjugate form. This would be different for active networks since non-conservative energy terms can arise from hermitian conjugate of either the system operators or the coupling operators in $H_{int}$ \citep{Gough_2010,Nurdin_2009_LQG,james2007hinfinity,James_2010}. A linear map in the spirit of Bogoliubov formulation is normally used to solve realization theory, and control theory for both passive and active networks with many quadratic degrees of freedom \citep{gough2013realization,Nurdin_2009_LQG}; however, we'll arrive at the same conclusion without taking in account this linear mapping.

\subsection{Transfer function}
In classical systems, the transfer function \textbf{H}$(s)$ (or transfer matrix in multiport networks) serves as a way to model characteristic solutions of time-invariant input-output problems in the Laplace domain \citep{Ogata_2010,Williams_2007}. Similarly, a transfer function for linear quantum networks with many (usually quadratic) degrees of freedom can be constructed from a state-space representation \citep{gough2013realization,grivopoulos2016transfer,Yanagisawa_2003_P1,Yanagisawa_2003_P2,yamamoto2016quantum,Zhang_2011,zhang2012response}  using Eq. (\ref{eq:ABCD_part_1})- (\ref{eq:ABCD_part_2}). This is done by  taking the Laplace transform of Eq. (\ref{eq:ABCD_part_1})-(\ref{eq:ABCD_part_2}), assuming that the initial state $a(0)$ in Laplace domain is zero, is 
\begin{align}
    b_{out}(s) & = (\mathbf{C}(s\mathbf{I}-\mathbf{A})^{-1} \mathbf{B} +  \mathbf{D} ) b_{in}(s),
\end{align}
where s is a complex variable, $\mathbf{I}$ is the identity matrix, and $(s\mathbf{I} - \textbf{A})$ is invertible. 

The transfer function is defined as a ratio of output over input in Laplace domain, $\mathbf{H}(s) = \frac{b_{out}(s)}{b_{in}(s)} = \mathbf{C}(s - \mathbf{A})^{-1} \mathbf{B} +  \mathbf{D}$. The transfer function for the lumped LC circuit results in 
\begin{align}
    \mathbf{H}(s) & = \frac{(s - \frac{\kappa^2}{2}) + i\triangle}{(s + \frac{\kappa^2}{2}) + i\triangle }
\end{align}
for Re$\{ s \} > 0$. For this derivation  we have treated \textit{ABCD} matrices and \textbf{H}$(s)$ as matrices when they are scalars in a 1-port network like the lumped LC problem we developed in this article. This notation is used to impart the idea that these quantities are matrices with dimensions corresponding to the number of input/output ports a network has. 

Throughout our derivation, we implicitly explained that the operators can be expressed as a Wiener noise probability distribution. This can be seen in Eq. (\ref{eq:bosonic_heinsenberg_part}), where the bosonic wave is dependent on the convolution of the system operator. This means the that we could have represented the system and bosonic dynamics using It$\Bar{o}$ calculus with the bosonic field as a quantum Wiener noise operator \citep{Gardiner_1985,Gardiner_1992}. 

\section{Conclusion}

In this article, we demonstrated how to construct  a simple  single-port quantum circuit from first principles of QION theory. A lumped LC circuit coupled to a transmission line can be easily modeled as an open quantum system, where the dynamical solutions can be turned into a state-space representation and a corresponding transfer function. We then provided a formalism of quantum circuits in terms of microwave engineering, resulting in an easier understanding to solving microwave systems. Our results imply that we can construct quantum networks based on fundamental theorems of QION, however, as we generalize to multi-port systems, QION modeling becomes increasingly difficult. Although \textit{SLH} theory is needed to accurately model multi-port networks, QION can do this for a simple single port.

\end{document}